\newif\ifShowKeys
\ShowKeystrue
\ShowKeysfalse

\documentclass[11pt,a4paper]{article} 
\pdfoutput=1
\usepackage[no-natbib-sort]{my-jheppub}
\usepackage{amsmath, amssymb}

\usepackage{bm}
\usepackage{environ}
\usepackage{mathrsfs}
\usepackage{array,arydshln}
\usepackage{booktabs}
\usepackage{subfigure}

\usepackage{graphicx,epsfig}
\usepackage{epic}

\usepackage{tensor} 							% Ratcliffe package to write tensors
\usepackage{youngtab}

\usepackage{float}
\usepackage[dvipsnames]{xcolor}
\definecolor{maroon}{rgb}{0.8,0.3,0.}

\usepackage{slashed}
\usepackage[nodayofweek]{date time}

\ifShowKeys \usepackage[notcite]{showkeys} \fi

\usepackage{hyperref}

\usepackage{aurical}
\usepackage[T1]{fontenc}

\usepackage{framed}
\definecolor{shadecolor}{RGB}{255, 230, 204}

\allowdisplaybreaks

\newmuskip\pFqmuskip

\newcommand*\pFq[6][8]{%
  \begingroup % only local assignments
  \pFqmuskip=#1mu\relax
  \mathcode`\,=\string"8000
  \begingroup\lccode`\~=`\,
  \lowercase{\endgroup\let~}\pFqcomma
  {}_{#2}F_{#3}{\left[\genfrac..{0pt}{}{#4}{#5};#6\right]}%
  \endgroup
}

\newcommand*\pFtildeq[6][8]{%
  \begingroup % only local assignments
  \pFqmuskip=#1mu\relax
  \mathcode`\,=\string"8000
  \begingroup\lccode`\~=`\,
  \lowercase{\endgroup\let~}\pFqcomma
  {}_{#2}\widetilde{F}_{#3}{\left[\genfrac..{0pt}{}{#4}{#5};#6\right]}%
  \endgroup
}

\newcommand{\pFqcomma}{\mskip\pFqmuskip}

\newcommand{\be}{\begin{equation}}
\newcommand{\ee}{\end{equation}}

\newcommand{\mc}{\mathcal }

\newcommand{\bv}{\begin{verbatim}}
\newcommand{\ev}{\end{verbatim}}

\newcommand{\la}{\label}

\newcommand{\eps}{\varepsilon}

\newcommand{\red}[1]{\textcolor{red}{#1}}

\title{Thermal properties of a string bit model at large $N$}

\author[a,b]{Matteo Beccaria}

\abstract{

We study the finite temperature properties of a recently introduced  string bit model
designed to capture some  features of the emergent string in 
the tensionless limit.
The model consists of a pair of bosonic and fermionic  bit operators transforming in the adjoint 
representation of 
the color group $SU(N)$.
Color confinement is not achieved as a dynamical effect, but instead is enforced by 
an explicit  singlet projection. At large $N$ and finite temperature, 
 the model has a non trivial thermodynamics.
In particular, there is a Hagedorn type transition 
at a finite temperature $T=T_{\rm H}$ where the string degrees of freedom are liberated and the 
free energy gets a large contribution $\sim N^{2}$ that plays  the role of an order parameter. 
For $T>T_{\rm H}$, the low temperature phase becomes unstable. In the new phase,
the thermodynamically
favoured configurations are characterized by a non-trivial gapped density of the 
$SU(N)$ angles associated with the singlet projection. We present an accurate algorithm for the 
determination of the density profile at $N=\infty$. In particular, we determine 
the gap endpoint at generic temperature and analytical expansions valid 
near the Hagedorn transition as well as at high temperature. The leading order corrections are 
characterized by non-trivial exponents that are determined analytically and compared with 
explicit numerical calculations.
}
\affiliation[a]{Dipartimento di Matematica e Fisica Ennio De Giorgi,\\
Universit\`a del Salento, Via Arnesano, 73100 Lecce, 
Italy} 
\affiliation[b]{Istituto Nazionale di Fisica Nucleare (INFN), Sezione di Lecce} 

\emailAdd{matteo.beccaria@le.infn.it} 

\vfill

\begin{document}

%\date{\currenttime}
%\begin{flushleft}\boxed{\small{\tt \today \ \ - \ \  \currenttime }}\end{flushleft}

%\begin{flushright}\small{Imperial-TP-AT-2017-{04}}\end{flushright}				% report number

\maketitle
\flushbottom

%\appendix
%\section{}\section{}\section{}\section{}\section{}\section{}\section{}
%\section{}\section{}\section{}

\section{Introduction and summary of  results}

Thorn's string bits models
have been originally proposed  as a description of 
superstrings where stability and causality are manifest 
 \cite{Giles:1977mpa,Thorn:1991fv,Bergman:1995wh,Sun:2014dga,Thorn:2014hia}. In the framework of 
't Hooft $1/N$ expansion and  light-cone parametrization of the string, 
one considers the continuum limit of very long chains composed of elementary string bits
transforming in the adjoint of the color gauge group $SU(N)$.
When the number  of bits $M$ gets large, the bit
chains behave approximately like continuous strings with recovered
Lorentz invariance.~\footnote{On general grounds, this requires also the number of colors
$N$ to be large. 
For recent numerical studies at finite $M, N$
see  \cite{Chen:2016hkz}.}
The finite temperature thermodynamics of such string bit models is quite rich in the 
't Hooft large $N$ limit. Stringy low energy states
 turn out to be  
color singlets separated from non-singlets by 
 an infinite gap in units of the characteristic singlet energy $\sim 1/M$  \cite{Sun:2014dga,Thorn:2014hia}.
 This means that color confinement emerges
 as a consequence of the dynamics. Besides, the singlet subspace exhibits a 
 Hagedorn transition \cite{Hagedorn:1965st} at infinite $N$ \cite{tHooft:1973alw,Thorn:1980iv}
 signalled by a divergence of the partition function for temperatures
 above a certain finite temperature $T>T_{\rm H}$. As usual, this behaviour is generically associated with 
 a density of states  growing exponentially with energy as in the original 
 dual resonance models \cite{Fubini:1969qb} or modern string theory  \cite{Atick:1988si}.
 When string perturbation theory is identified with the 't Hooft $1/N$ expansion of string bit dynamics,
 the $N=\infty$ Hagedorn transition is consistent the interpretation of $T_{\rm H}$ in the free string 
 as an artifact of the  zero coupling limit \cite{Atick:1988si} with a possible phase transition near 
 $T_{\rm H}$ to a  phase dominated by the fundamental degrees of freedom of the emergent
 string theory. 
 
Recently and remarkably,
the Hagedorn transition of string bit models
has been further clarified \cite{Thorn:2015bia}, and 
discovered  also in simpler reduced systems where the singlet restriction is imposed
from the beginning as a kinematical constraint and not as a dynamical feature  
\cite{Raha:2017jgv,Curtright:2017pfq}.
The starting point is  the thermal partition function 
\be
\la{1.1}
Z = \text{tr}\,e^{-\beta\,(H+\mu\,M)},
\ee
where $\beta$ is the inverse temperature, $H$ the string bit model Hamiltonian, and $M$ is the bit number operator associated
with the chemical potential $\mu$.
The partition function (\ref{1.1}) is quite natural and has a simple origin from  the light-cone description of the 
emergent string  where $H = P^{-}/\sqrt 2$,  $\mu\,M = P^{+}/\sqrt{2}$, 
and thus $H+\mu\,M = P^{0} = (P^{+}+P^{-})/\sqrt 2$ \cite{Goddard:1973qh}. 
The reduced model considered in 
 \cite{Raha:2017jgv,Curtright:2017pfq} is the projection of (\ref{1.1}) 
  on the subspace of singlets states with  $H=0$, {\em i.e.} for the associated tensionless string, 
   and is described by  the simpler partition function 
%  \footnote{The group theoretical meaning of the coefficient of $x^{n}$ in $Z_{0}$ 
%is the  number of color singlets 
%in the  symmetrized tensor products of adjoints (bosonic states) 
%times the antisymmetrized tensor products of adjoints (fermionic states). The integer exponent $n$
%is the total number of adjoint factors.}
\be
\la{1.2}
Z_{0}= \text{tr}_{\rm singlets}\,x^{M}, \qquad x = e^{-\beta\,\mu}.
\ee
Here, we shall focus on the simple model considered in \cite{Raha:2017jgv} which consists of one
pair of bosonic and
fermionic string bits operators $a$ and $b$, both  transforming in the adjoint of $SU(N)$. \footnote{
Before singlet projection, the large $N$ limit of the string bit model describes a 
non-covariant subcritical light-cone string with no transverse coordinates and one 
Grassmann world-sheet field. In general, an important feature of 
string bit models is that they can be formulated in a space-less
fashion with emerging spatial transverse and longitudinal coordinates \cite{Thorn:2014hia}. Thus,
they
may be regarded as a realization of 't Hooft holography \cite{tHooft:1993dmi}.
}
Extensions to  models with more bit species and discussion of $1/N$ corrections have been 
addressed in \cite{Curtright:2017pfq}. The bit number operator
is $M = \text{tr}(\overline a\,a+\overline b\,b)$, where trace is in color space, 
 and the projected partition function  (\ref{1.2})
can be computed by  group averaging according to the analysis of 
\cite{Raha:2017jgv,Curtright:2017pfq} \footnote{
The prefactor $(1-x)/(1+x)$ in (\ref{1.3}) takes into account that 
the bit operators $a,b$ are traceless  and hence are adjoints under $SU(N)$. 
}
\begin{align}
\la{1.3}
Z_{0} &= \frac{1-x}{1+x}\,\int dU(\bm{\vartheta})
\,\text{tr} (x^{M}\,e^{i\,G_{k}\vartheta_{k}}) = 
\frac{1-x}{1+x}\,\int dU(\bm{\vartheta})
\,\prod_{1\le k < \ell\le N}\frac{1+x\,e^{i\,(\vartheta_{k}-\vartheta_{\ell})}}
{1-x\,e^{i\,(\vartheta_{k}-\vartheta_{\ell})}} \notag \\
&= \left(\frac{1+x}{1-x}\right)^{N-1}\,\int dU(\bm{\vartheta})
\,\prod_{1\le k < \ell\le N}\frac{1+x^{2}+2\,x\,\cos(\vartheta_{k}-\vartheta_{\ell})}
{1+x^{2}-2\,x\,\cos(\vartheta_{k}-\vartheta_{\ell})},
\end{align}
where $G_{k}$ span the Cartan subalgebra of $U(N)$. 
The group integration in (\ref{1.3})
is with respect to the normalized Haar measure
\be
dU(\bm{\vartheta}) =  \frac{1}{N!\,(2\pi)^{N}}\,\int_{-\pi}^{\pi}d^{N}\bm{\vartheta}\,\prod_{1\le k < \ell \le N}
\,4\,\sin^{2}\left(\frac{\vartheta_{k}-\vartheta_{\ell}}{2}\right).
\ee
In the 't Hooft large $N$ limit, the partition function 
 (\ref{1.3}) may be evaluated by saddle point methods. The dominant saddle contribution
 is characterized by a continuous density
of phases $\rho(\vartheta; x)$.
The analysis of  \cite{Raha:2017jgv,Curtright:2017pfq} shows that there exists, for $N=\infty$, a 
critical point $x_{\rm H}=1/2$. For low temperatures $x<x_{\rm H}$, 
 the stable solution of the saddle point condition is 
  associated with a uniform constant density $\rho(\vartheta; x) = 1/(2\pi)$
and a partition function that has a  finite $N\to \infty$ limit.
Instead, above the Hagedorn temperature, {\em i.e.} for $x>x_{\rm H}$, the density $\rho(\vartheta; x)$ 
is a non trivial function which is non zero on a finite subinterval
 $|\vartheta|\le\vartheta_{0}(x)<\pi$.
 In this gapped phase, the partition function has the leading large $N$ behaviour 
$\log Z_{0}  = N^{2}\,F_{2}(x)
+\mc O(N\,\log N)$ where $F_{2}(x)$ is  a function of the temperature  growing
monotonically  from $F_{2}(1/2)=0$ up to 
$F_{2}(1) = \log 2$. This function may be regarded as an order parameter that 
measures the smooth activation of the 
string bit degrees of freedom above the Hagedorn temperature.

This change of behaviour at $x=x_{\rm H}$
is similar to what happens in the unitary matrix model transition \cite{Gross:1980he}
with the  coupling constant of the latter being traded here by the temperature parameter $x$. 
Similar results have also been obtained in \cite{Aharony:2003sx} for free adjoint $U(N)$
SYM on $S^{3}\times \mathbb{R}$, see also \cite{Sundborg:1999ue}. More generally,
in the context of AdS/CFT
duality, it is an important issue to  understand the thermodynamics of specific conformal theories
with singlet constraint, see for instance   
\cite{Skagerstam:1983gv,Aharony:2003sx,Schnitzer:2004qt,
Schnitzer:2006xz,Shenker:2011zf} and the recent  M-theory  motivated study \cite{Beccaria:2017aqc}.

At temperatures above the Hagedorn
transition, the precise form of the phase density profile $\rho(\vartheta; x)$ is 
not known in analytic form, not even in the strict $N=\infty$ limit. 
The aim of this paper is to provide more information 
about this quantity and the related width $\vartheta_{0}(x)$. 

To this aim, following the strategy of 
\cite{Aharony:2003sx}, we reconsider the solution
of the partition function for $U(N)$ gauge theory on a 2d lattice at large $N$ for a broad class of 
single-plaquette actions found in \cite{Jurkiewicz:1982iz}. We exploit it in order to  cast the homogenous
integral equation governing $\rho(\vartheta; x)$ into an infinite dimensional linear system involving 
the higher (trigonometric) 
momenta of $\rho$. Truncation to a finite number of modes provides an accurate algorithm
for the determination of the density. As we shall discuss, 
the outcome is not only numerical because
 some analytical information
can be extracted from the above mentioned linear system. Besides, analysis of the  numerical  data produced by the algorithm
suggests how to extract precise analytical information from the integral equation in certain limits. 
A summary of our results follows:
\begin{enumerate}
\item For $x\to x_{\rm H}= 1/2$ the distribution 
gap closes, {\em i.e.} $\vartheta_{0}(x)\to \pi$, with a correction
vanishing as
$\sim (T-T_{\rm H})^{1/4}$, Near $x_{\rm H}$, the phase density approaches
a Wigner semicircle law (in the variable $\sin(\vartheta/2)$).  
\begin{align}
\la{1.5}
\vartheta_{0}(x) &= \pi
-2\,\sqrt{2}\,\left(\frac{x-x_{\rm H}}{2}\right)^{1/4}-\frac{2\sqrt{2}}{3}\,\left(
\frac{x-x_{\rm H}}{2}\right)^{3/4}
+\dotsb, \notag \\
\rho(\vartheta; x\to x_{\rm H}) &\sim
\frac{1}{\pi\,\sin^{2}(\vartheta_{0}/2)}\,
\,\left(\sin^{2}\frac{\vartheta_{0}}{2}-\sin^{2}\frac{\vartheta}{2}\right)^{1/2}\,
\cos\frac{\vartheta}{2}.
\end{align}

\item At high temperature, $x\to 1$, the phase distribution collapses with 
 $\vartheta_{0}(x)\sim T^{-1/3}$. A non uniform quadratic distribution is achieved inside $[-\vartheta_{0}, \vartheta_{0}]$
\begin{align}
\la{1.6}
\vartheta_{0}(x\to 1) &= \left[6\,\pi\,(1-x)\right]^{1/3}+\dotsb, \notag \\
\rho(\vartheta; x\to 1) &\sim \frac{3}{4\,\vartheta_{0}^{3}}\,\left(\vartheta_{0}^{2}-\vartheta^{2}\right).
\end{align}

\item The order parameter, {\em i.e.} the function $F_{2}(x)$ appearing in the expansion
 $\log Z_{0} = N^{2}\,F_{2}(x) + \dotsb$, admits the following expansions around $x=x_{\rm H}$
 and $x=1$
\begin{align}
\la{1.7}
F_{2}(x\to x_{\rm H}) = \frac{1}{2}\,(x-x_{\rm H})+\dotsb,
\qquad
F_{2}(x\to 1) = \log 2-\frac{3\,(6\pi)^{2/3}}{20}\,(1-x)^{2/3}+\dotsb.
\end{align}
The first expansion shows that $F_{2}(x)$ is linear just above $x_{\rm H}$ as originally
suggested in  \cite{Thorn:2015bia}.  The second expansion shows the leading
correction to the known infinite temperature limit $\log 2$.
\end{enumerate}

The plan of the paper is the following. In Sec.~(\ref{sec:basic}) we present the integral
equation for the phase density $\rho(\vartheta; x)$ discussing first some of its features at finite $N$.
Then, our proposed $N=\infty$ self-consistent algorithm and its predictions are presented.
Sec.~(\ref{sec:exp})
is devoted to the derivation of various analytical expansions. In particular, in 
Sec.~(\ref{sec:exp-hag}) and (\ref{sec:exp-high}) we discuss the expansion of the phase 
density near the Hagedorn temperature and at high temperature $x\to 1$. The 
behaviour of the partition function near $x=x_{\rm H}$ and $x=1$
is considered in Sec.~(\ref{sec:exp-logZ}).
Conclusions and open directions are briefly discussed in a final section.

\section{Self-consistent determination of the density at $N=\infty$}
\la{sec:basic}

%\subsection{Finite $N$ data \red{TOGLIERE sec ? }}
%\la{sec:finiteN}

\begin{figure}[t]  % !htb]
\centering
\subfigure{
\includegraphics[scale=0.3]{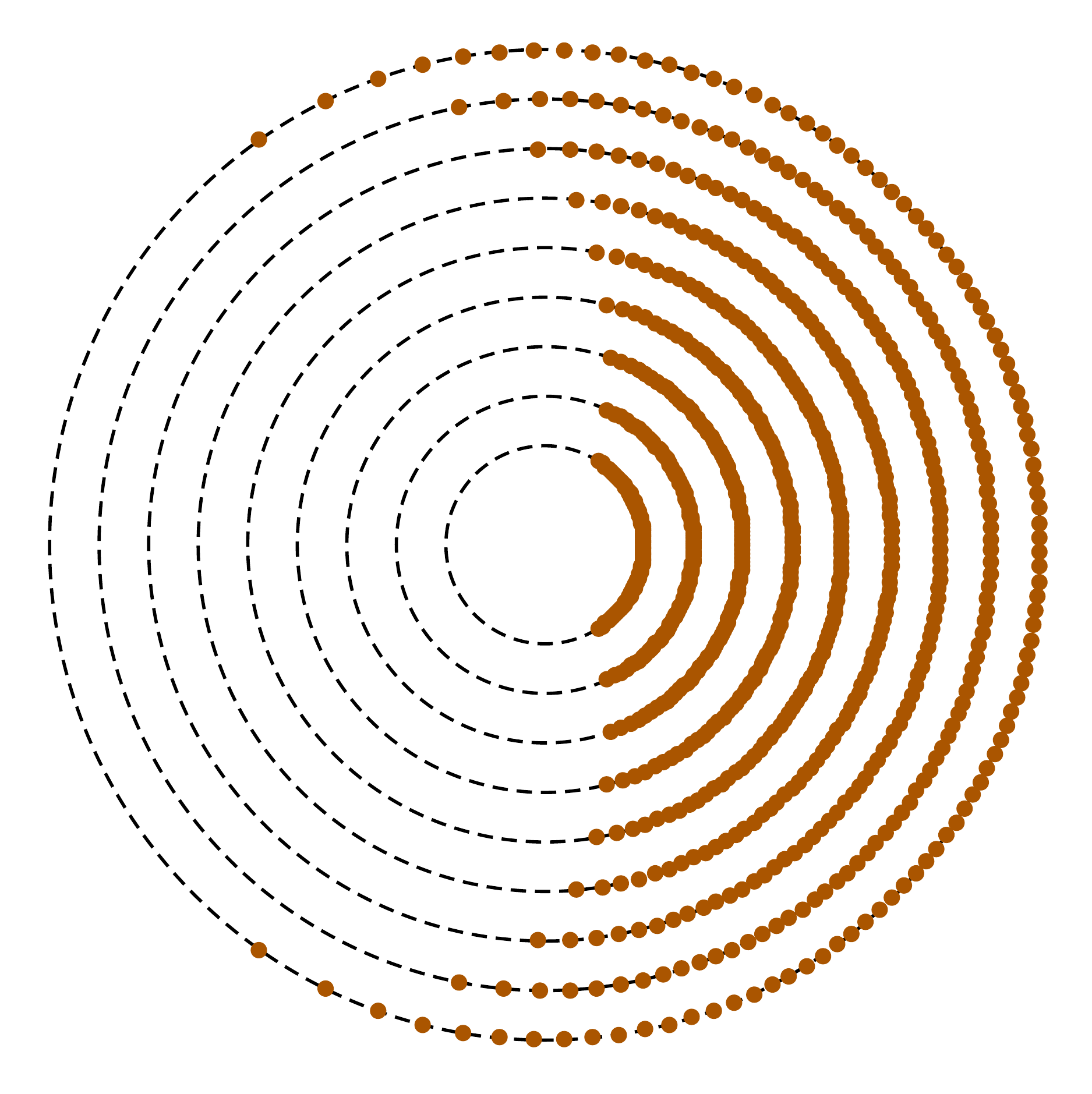}}
\subfigure{
\includegraphics[scale=0.4]{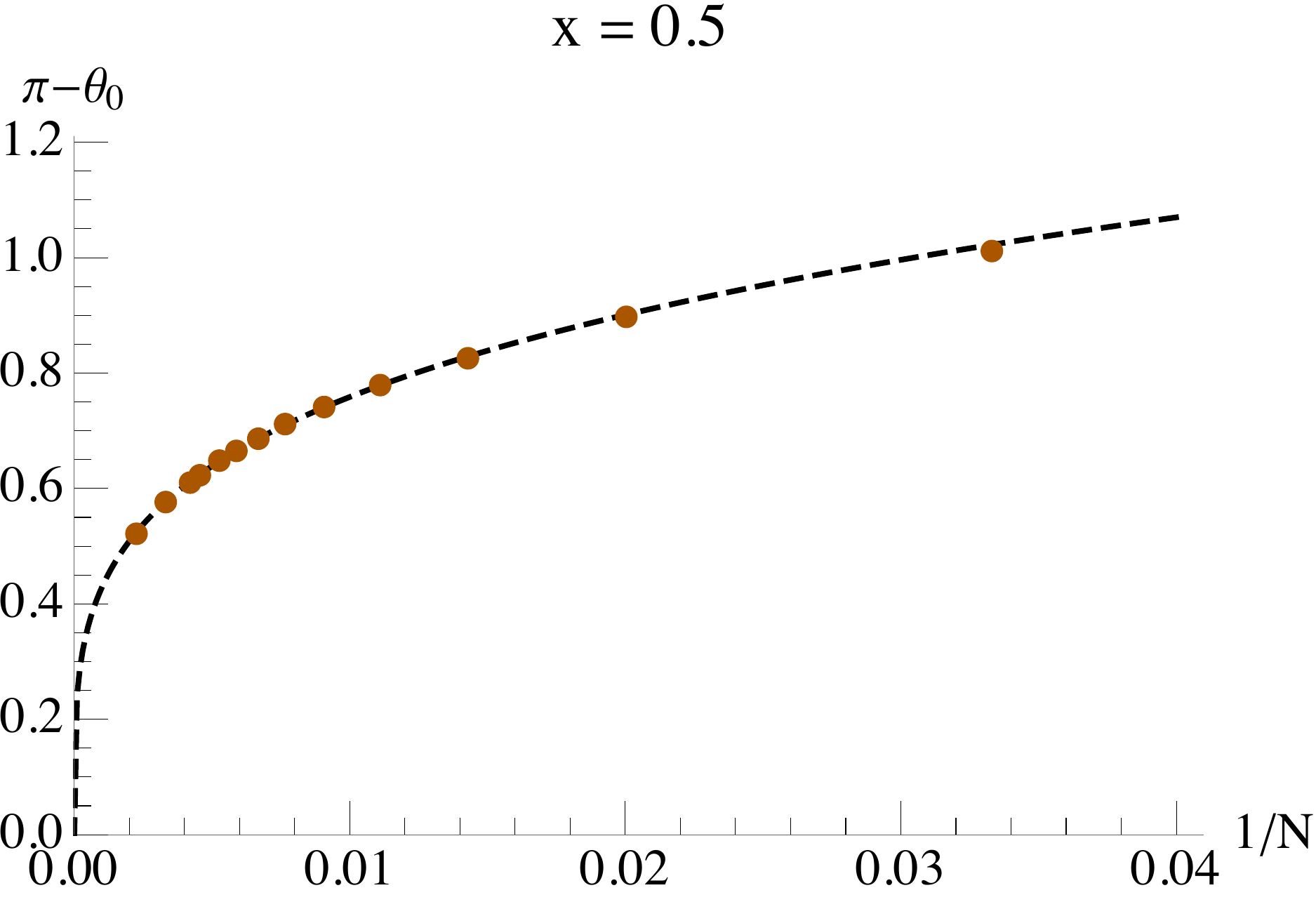}}
\caption[] {
Finite $N$ solution of the discrete phase equation (\ref{2.1}). {\bf Left:} roots of (\ref{2.1}) for $N=100$
drawn as phases on circles of different radii. The outmost circle has $x=0.5$, while the inner
circles have $x$ increased in steps $0.05$ up to $x=0.9$. {\bf Right:} gap half-width $\pi-\vartheta_{0}$
evaluated at $x=x_{\rm H}$ with increasing $N$ up to 450. The dashed line is a power law fit 
providing an exponent very close to $1/4$, {\em i.e.} $\vartheta_{0}(x_{\rm H})=\pi+\mc O(N^{-1/4})$.
%It is clear from this figure that very large $N$ are needed to
%accurately describe the $N\to \infty$ limit. 
\label{fig:finiteN}}
\end{figure}

As discussed in \cite{Raha:2017jgv}, the determination of the saddle point $\bm{\vartheta}$ of
(\ref{1.3}) for finite $N$ 
amounts to finding the solution of the set of equations
\be
\la{2.1}
\sum_{\ell\neq k}\cot\left(\frac{\vartheta_{k}-\vartheta_{\ell}}{2}\right)-\frac{4\,x\,(1+x^{2})\,\sin(
\vartheta_{k}-\vartheta_{\ell})}{1+x^{4}-2\,x^{2}\,\cos(2\,(\vartheta_{k}-\vartheta_{\ell}))}=0.
\ee
The numerical solution of (\ref{2.1}) for $N=100$ and $x>x_{\rm H}$ is shown in the left panel of 
Fig.~(\ref{fig:finiteN})
where one appreciates the opening of a gap whose width increases as $x\to 1$. The distribution of the 
roots $\bm{\vartheta}$ is non trivial, {\em i.e.} it is not uniform. 
Precisely at the $N=\infty$ Hagedorn transition point, $x=x_{\rm H}$, the 
gap closes as $N\to \infty$ according to  the  finite size scaling 
$\pi-\vartheta_{0} = \mc O\left(N^{-\delta}\right)$ with $\delta\simeq 1/4$, as shown in the right panel.
This slow convergence of observables at increasing $N$ means that a reliable characterization of the model  for $N=\infty$
is difficult by extrapolation from finite $N$ data. Besides, we are interested in analytical expansions
near Hagedorn transition as well as at high temperature. For these reasons, we present in the next 
section a self-consistent accurate treatment of the $N=\infty$ limit that will prove itself to be more effective
than finite $N$ extrapolation.

%\subsection{Self-consistent solution at $N=\infty$}
%\la{sec:self}

\bigskip

At $N\to \infty$, the roots of (\ref{2.1}) are described by a smooth density 
$\rho(\vartheta; x)$ which is positive for $|\vartheta|<\vartheta_{0}(x)$ and vanishes at $
\vartheta=\pm \vartheta_{0}$.
Taking the continuum limit of (\ref{2.1}), the function
$\rho(\vartheta; x)$ obeys the homogeneous integral equation
\begin{align}
\la{2.2}
& \int_{-\vartheta_{0}(x)}^{\vartheta_{0}(x)} d\vartheta\,G(\vartheta'-\vartheta; x)\,
\rho(\vartheta; x) = 0, \notag \\
& G(\vartheta; x) = \cot\left(\frac{\vartheta}{2}\right)-\frac{4\,x\,(1+x^{2})\,\sin\vartheta}{x^{4}+1
-2\,x^{2}\,\cos(2\vartheta)}.
\end{align}
To solve it, we  exploit the remarkably simple identity
\be
\la{2.3}
\frac{x\,(1+x^{2})\,\sin\vartheta}{x^{4}+1
-2\,x^{2}\,\cos(2\vartheta)} = \sum_{n=0}^{\infty}
x^{2n+1}\,\sin((2n+1)\,\vartheta),
\ee
that holds in our case, {\em i.e.} for  $0<x<1$ and real $\vartheta$. The expansion (\ref{2.3}) allows to 
 write  (\ref{2.2}) in the form 
\begin{align}
\la{2.4}
& \int_{-\vartheta_{0}(x)}^{\vartheta_{0}(x)}d\vartheta\,\cot\left(\frac{\vartheta'-\vartheta}{2}\right)\,\rho(\vartheta; x) =\notag \\ 
& \qquad \qquad 4\,\sum_{n=0}^{\infty}
x^{2n+1}\,\int_{-\vartheta_{0}(x)}^{\vartheta_{0}(x)}d\vartheta\,\sin((2n+1)\,(\vartheta'-\vartheta))\,\rho(\vartheta; x).
\end{align}
Taking into account that the density is expected to be even, $\rho(\vartheta; x) = \rho(-\vartheta; x)$, 
we can further simplify (\ref{2.4}) and obtain 
\begin{align}
\la{2.5}
& \int_{-\vartheta_{0}(x)}^{\vartheta_{0}(x)}d\vartheta\,\cot\left(\frac{\vartheta'-\vartheta}{2}\right)\,
\rho(\vartheta; x) = \notag \\
& \qquad \qquad 4\,\sum_{n=0}^{\infty}
x^{2n+1}\,\sin((2n+1)\vartheta')\,\int_{-\vartheta_{0}(x)}^{\vartheta_{0}(x)}
d\vartheta\,\cos((2n+1)\,\vartheta)\,\rho(\vartheta; x).
\end{align}
It is convenient to recast (\ref{2.5}) in the apparently inhomogeneous form  
\be
\la{2.6}
\int_{-\vartheta_{0}(x)}^{\vartheta_{0}(x)}d\vartheta\,\cot\left(\frac{\vartheta'-\vartheta}{2}\right)\,\rho(\vartheta) = 
4\,\sum_{n=0}^{\infty}\rho_{n}\,
x^{2n+1}\,\sin((2n+1)\vartheta'),
\ee
where we have introduced the trigonometric momenta
\be
\la{2.7}
\rho_{n}(x) = \int_{-\vartheta_{0}(x)}^{\vartheta_{0}(x)}d\vartheta\,
\cos((2n+1)\,\vartheta)\,\rho(\vartheta;x).
\ee
As discussed in \cite{Aharony:2003sx}, the general solution of the problem (\ref{2.6}) is known and 
reads \footnote{
%The original application of the solution (\ref{2.8}) was for a given source term in the r.h.s. of 
%(\ref{2.6}). 
A self-consistent interpretation of the solution (\ref{2.8}) first appeared in 
\cite{Aharony:2003sx} in a different context, see also the recent application \cite{Beccaria:2017aqc}.
}
\begin{align}
\la{2.8}
\rho(\vartheta) &= \frac{1}{\pi}\sqrt{\sin^{2}\left(\frac{\vartheta_{0}}{2}\right)
-\sin^{2}\left(\frac{\vartheta}{2}\right)}\,\sum_{m=1}^{\infty}Q_{m}\,\cos\left[
(m-\tfrac{1}{2})\,\vartheta\right],
\notag \\
Q_{m} &= \mathop{\sum_{\ell=0}^{\infty}}_{\frac{m+\ell-1}{2} = 0, 1, 2, \dots}4\,x^{m+\ell}\rho_{\frac{m+\ell-1}{2}}\,P_{\ell}(\cos\vartheta_{0}),
\end{align}
where, for brevity, we have omitted the explicit dependence on $x$. 
%The gap width $\vartheta_{0}$ is determined by 
%\be
%Q_{1} = Q_{0}+2.
%\ee
We can now truncate the expansion (\ref{2.8})  by keeping only 
a fixed number of terms $\bm{\rho}^{(K)} = \{\rho_{k}\}_{k=0,\dots, K}$.
The density is thus written in terms of the finite set of quantities $\bm{\rho}^{(K)}$. 
Replacing the density  expression into (\ref{2.7}) we obtain a homogeneous linear system
\be
\la{2.9}
\mc M^{(K)}(x, \vartheta_{0}^{(K)})\,\bm{\rho}^{(K)}=0.
\ee
Non trivial solutions exists only if 
\be
\la{2.10}
\det\mc M^{(K)}(x, \vartheta_{0}^{(K)})=0,
\ee
which is the condition that 
determines the approximate gap width $\vartheta_{0}^{(K)}$ for each $x>x_{\rm H}$.
Once  $\vartheta_{0}^{(K)}$ is computed, we solve (\ref{2.9}) for the eigenvector $\bm{\rho}^{(K)}$ and 
obtain the density from (\ref{2.8}). The eigenvector normalization is fixed by requiring
$\rho(\vartheta)$ to be normalized with unit integral. To appreciate the accuracy of the method,
we show in Tab.~(\ref{tab1}) the solution $\vartheta_{0}^{(K)}(x)$ of (\ref{2.10})
evaluated at various $x>x_{\rm H}$, and with $K$ growing 
from 10 to 34. 
%
%\begin{table}[H]
%\be
%\def\arraystretch{1.3}
%\begin{array}{cccccc}
%\toprule
%K &\phantom{xx}& \vartheta_{0}^{(K)}(0.6) &  \vartheta_{0}^{(K)}(0.7)&  \vartheta_{0}^{(K)}(0.8)&  \vartheta_{0}^{(K)}(0.9) \\
%\midrule
%10 && 0.75611798 & 0.67837211 & 0.61735541 & 0.52688066 \\
% 14 && 0.75611796 & 0.67837140 & 0.61733829 & 0.52675452 \\
% 18 && 0.75611796 & 0.67837138 & 0.61733787 & 0.52672285 \\
% 22 && 0.75611796 & 0.67837137 & 0.61733775 & 0.52671194 \\
% 26 && 0.75611796 & 0.67837137 & 0.61733773 & 0.52670810 \\
% 30 && {\bf 0.75611796} & {\bf 0.67837137} & {\bf 0.61733773} & {\bf 0.52670}691 \\
%\bottomrule
%\end{array}\notag
%\ee
%%%%%%%%%%%%%%%
%\caption{Solution of the condition (\ref{2.10}) for various $x>x_{\rm H}$ and increasing number of 
%modes $K$. As a guide, we write in bold face the digits in the last line that have not
%changed upon increasing $K$ from 26 to 30.}
%\label{tab1}
%\end{table}
%
\begin{table}[H]
\be
\def\arraystretch{1.3}
\begin{array}{cccccc}
\toprule
K &\phantom{xx}& \vartheta_{0}^{(K)}(0.6) &  \vartheta_{0}^{(K)}(0.7)&  \vartheta_{0}^{(K)}(0.8)&  \vartheta_{0}^{(K)}(0.9) \\
\midrule
10 && 0.7561179\red{8} & 0.67837\red{211} & 0.6173\red{5541} & 0.526\red{88066} \\
 14 && 0.75611796          & 0.678371\red{40} & 0.61733\red{829} & 0.5267\red{5452} \\
 18 && 0.75611796          & 0.6783713\red{8} & 0.617337\red{87} & 0.5267\red{2285} \\
 22 && 0.75611796          & 0.67837137           & 0.6173377\red{5} & 0.5267\red{1194} \\
 26 && 0.75611796          & 0.67837137           & 0.61733773 & 0.52670\red{810} \\
 30 && 0.75611796          &  0.67837137          & 0.61733773 &  0.526706\red{91} \\
 34 && 0.75611796          &  0.67837137           & 0.61733773 & 0.52670659 \\ 
\bottomrule
\end{array}\notag
\ee
%%%%%%%%%%%%%%
\caption{Solution of the condition (\ref{2.10}) for various $x>x_{\rm H}$ and increasing number of 
modes $K$. As a guide, we write in red the digits that change moving to the next row.}
\label{tab1}
\end{table}
The convergence appears to be exponential in $K$ although with a decreasing rate as 
$x\to 1$. This is because the
effect of the convergence factors $x^{2n+1}$ in (\ref{2.4}) is reduced. Nevertheless, still at $x=0.9$, 
the accuracy is of about 6 digits for $K=34$.

\begin{figure}[t]  % !htb]
\centering
\includegraphics[scale=0.5]{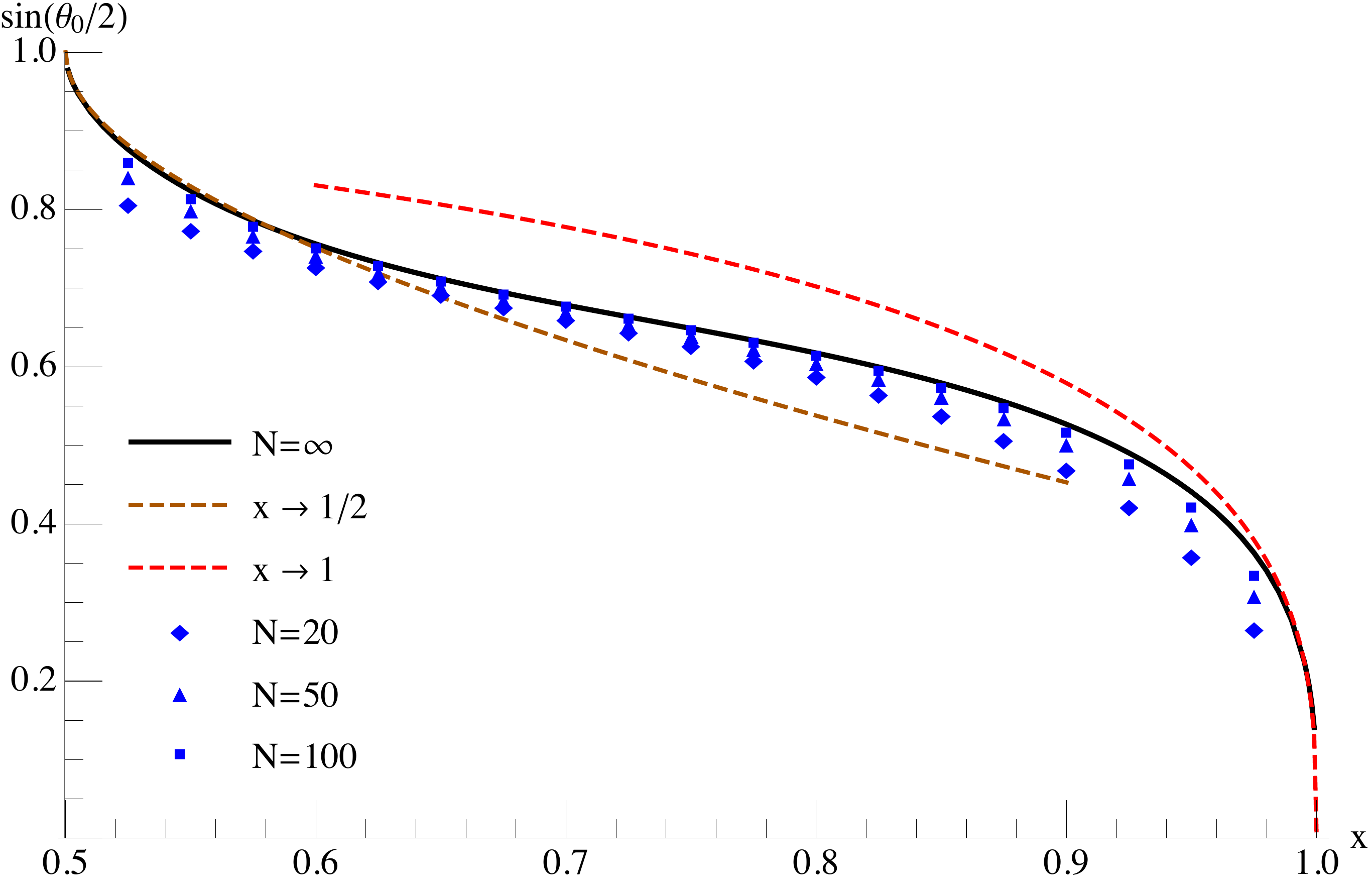}
\caption[] {
Temperature dependence of the phase gap $\vartheta_{0}(x)$. The black central curve is the result of 
the $N=\infty$ algorithm keeping $K=34$ modes. The blue symbols show the finite $N$ results 
at $N=20, 50, 100$ from the solution of (\ref{2.1}). The dashed brown and red curves are the analytical
approximations in (\ref{2.11}).
\label{fig:results}}
\end{figure}

\begin{figure}[t]  % !htb]
\centering
\subfigure{
\includegraphics[scale=0.33]{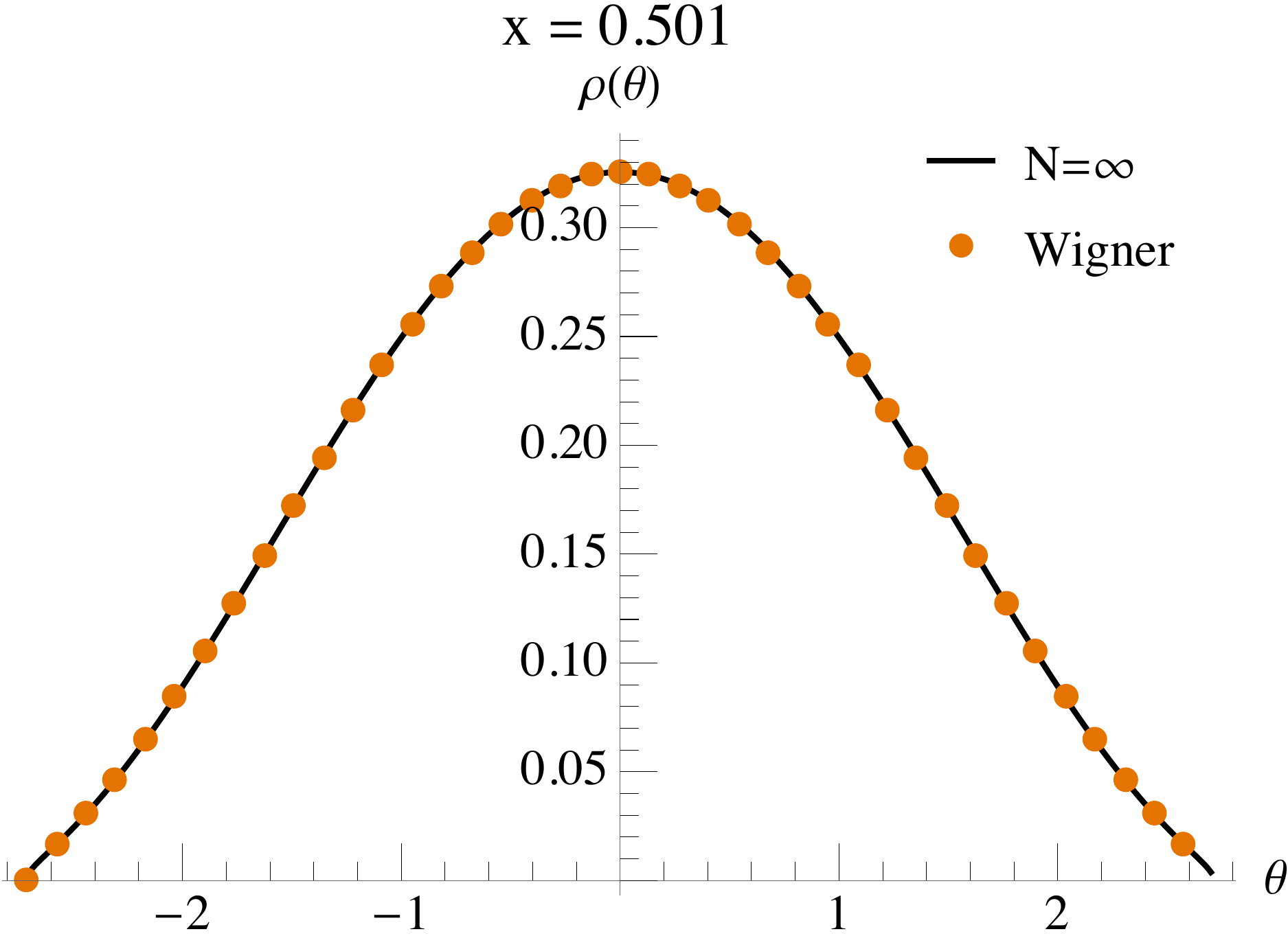}}
\subfigure{
\includegraphics[scale=0.33]{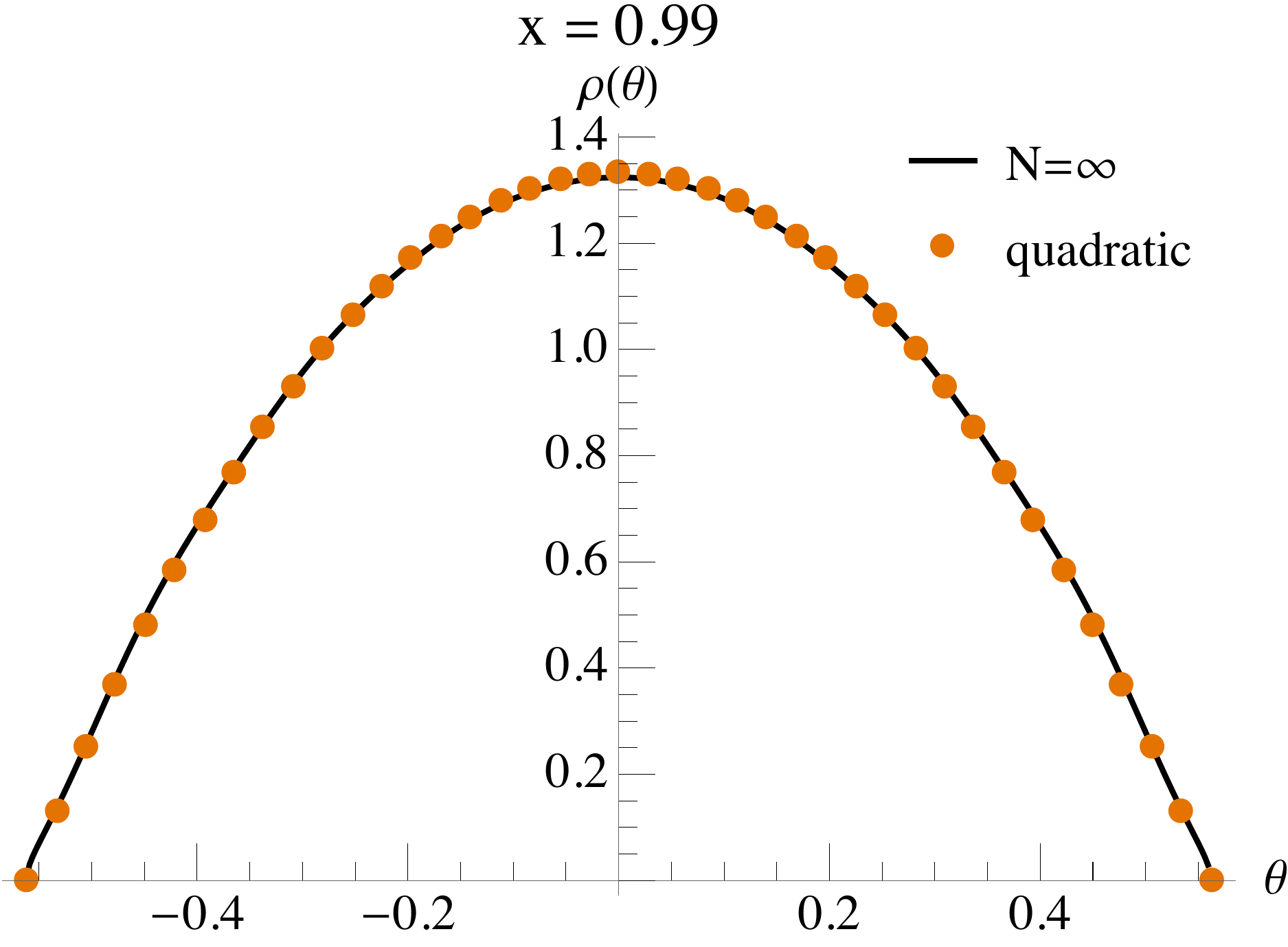}}
\caption[] {
Temperature dependence of the phase density $\rho(\vartheta; x)$. {\bf Left:} Just above the Hagedorn
transition. The black line is the density obtained by plugging in (\ref{2.8}) the solution of (\ref{2.10}).
The orange points are sample evaluations of (\ref{2.12}) and superimpose quite well. {\bf Right:} 
Near $x=1$. Again, the black line is the result from the self-consistent algorithm, while 
the orange points are samples of (\ref{2.13}). Apart from the very ends of the distribution, the agreement
is very good.
\label{fig:rho}}
\end{figure}

Working out the prediction of the above algorithm in the interval $x_{\rm H}< x <1$ we obtain
the black curve in Fig.~(\ref{fig:results}) where we plot $\sin(\vartheta_{0}(x)/2)$ vs. $x$. To appreciate
the convergence with $N$, we also 
show some  sample points obtained at finite $N=20,50,100$  from the solution of (\ref{2.1}). The dashed
curves are  analytical approximations valid around $x_{\rm H}$ and $x=1$ derived in the next
section, {\em i.e.} \footnote{The expansion of $\vartheta_{0}(x)$ in (\ref{2.11}) is 
an equivalent form of  (\ref{1.5}).}
\begin{align}
\la{2.11}
x &\to x_{\rm H}:\qquad 
\sin\frac{\vartheta_{0}(x)}{2} = 1-\sqrt\frac{x-\frac{1}{2}}{2}-\frac{1}{4}\left(x-\frac{1}{2}\right)
+\dotsb, 
\notag \\
x &\to 1:\qquad \vartheta_{0}(x) = \left[6\,\pi\,(1-x)\right]^{1/3}+\dotsb.
\end{align}
As we shall discuss later, the self-consistent determination of $\rho(\vartheta)$ provides 
also analytical information near the Hagedorn transition. We shall see that only the first
term in (\ref{2.8}) survives. This shows that  $\rho(\vartheta)$ is well described by 
\be
\la{2.12}
x \to x_{\rm H}:\qquad \rho(\vartheta; x) \to \frac{1}{\pi\,\sin^{2}(\vartheta_{0}/2)}\,\left(\sin^{2}\frac{\vartheta_{0}}{2}-\sin^{2}\frac{\vartheta}{2}\right)^{1/2}\,
\cos\frac{\vartheta}{2},
\ee
which is Wigner semi-circle law in the variable $\sin(\vartheta/2)$, well known in the theory of 
random symmetric matrices. Strictly at $x=x_{\rm H}$ this 
reduces to $\rho(\vartheta; 1/2) = \frac{1}{\pi}\,\cos^{2}\frac{\vartheta}{2}$. For $x\to 1$, 
we have found that the phase density is very well described by a quadratic law inside its support, {\em i.e.}
\be
\la{2.13}
x \to 1:\qquad \rho(\vartheta; x) \to \frac{3}{4\,\vartheta_{0}^{3}}\,(\vartheta_{0}^{2}-\vartheta^{2}),
\ee
as will also be confirmed analytically in the next section. The limiting forms (\ref{2.12}) 
and (\ref{2.13}) are tested in Fig.~(\ref{fig:rho}). In the two panels, we show the exact
density profile from the self-consistent algorithm and the predictions (\ref{2.12}) and (\ref{2.13}) at $x=0.501$ and $x=0.99$ 
respectively. The horizontal scale in the two panels is quite different due to the wide variation of 
$\vartheta_{0}(x)$. Up to a rescaling, the gross
shape of the two densities is roughly similar, although the two regimes 
 are clearly associated with different 
functions (semi-circle and quadratic).

\section{Analytical expansions}
\la{sec:exp}

In this section, we derive the analytical expansions (\ref{2.10}) characterizing the phase density 
$\rho(\vartheta;x)$  and its endpoint $\vartheta_{0}(x)$ near the Hagedorn transition and at
very high temperature $x\to 1$. 

\subsection{Opening of the gap near the Hagedorn transition}
\la{sec:exp-hag}

The condition (\ref{2.10}) may be solved perturbatively around $x=x_{\rm H}$. It is an algebraic 
equation in the variables $x$ and $h = \sin(\vartheta_{0}^{(K)}/2)$ whose
 complexity increases rapidly with $K$.
Just to give an example, for the almost trivial case $K=1$ we have the constraint
\begin{align}
K=1:\quad & 1+2 h^2 \left(h^2-2\right) x = 0.
\end{align}
The branch starting at $(x,h) = (1/2,1)$ has the expansion
\be
K=1:\quad h = 1-\left(\frac{x-x_{\rm H}}{2}\right)^{1/2}-\frac{1}{4}\,(x-x_{\rm H})
+\frac{3}{2}\left(\frac{x-x_{\rm H}}{2}\right)^{3/2}+\dotsb.
\ee
For $K=2$, the condition (\ref{2.10}) is much more complicated and reads
\begin{align}
K=2:\quad & 1+2 h^2 \left(h^2-2\right) x+2 h^2 \left(100 h^{10}-312 h^8+366 h^6-200 h^4+51 h^2-6\right) x^3\notag \\
&+4 h^8
   \left(25 h^8-152 h^6+288 h^4-224 h^2+64\right) x^4 = 0.
\end{align}
Expanding again around $x_{\rm H}$ we find 
\be
K=2:\quad h = 1-\left(\frac{x-x_{\rm H}}{2}\right)^{1/2}-\frac{1}{4}\,(x-x_{\rm H})
-\frac{33}{2}\left(\frac{x-x_{\rm H}}{2}\right)^{3/2}+\dotsb.
\ee
Repeating the procedure for increasing $K$, one finds that the first two terms of the expansion 
of $h$ are independent on $K$,
\be
h = 1-\left(\frac{x-x_{\rm H}}{2}\right)^{1/2}-\frac{1}{4}\,(x-x_{\rm H})
+\text{c}^{(K)}\,\left(\frac{x-x_{\rm H}}{2}\right)^{3/2}+\dotsb.
\ee
while the values of the third coefficient are 
\be
\la{3.6}
c^{(K)} = \frac{3}{2},-\frac{33}{2},-\frac{199}{6},-\frac{793}{18},-\frac{76153}{1530},-\frac{2484163}{4743
   0},-\frac{5915131}{110670},-\frac{32551537891}{604368870},\dotsb.
\ee
Increasing $K$ up to 30 and working with exact rational values, 
this sequence converges numerically to an asymptotic value that can be estimated by Wynn 
acceleration algorithm \cite{bender2013advanced}. The results are quite stable and independent on the Wynn algorithm 
parameter and give  $c^{(\infty)} = -54.0888227$. Such a large value suggests that the expansion (\ref{3.6}) could be 
only asymptotic, as expected near a phase transition.

Plugging the expansion (\ref{3.6}) in the linear system (\ref{2.9}) one finds that all $\rho_{n>0}$
vanish linearly with $x-x_{\rm H}$. This leaves the semi-circle asymptotic density that we wrote in
(\ref{2.12}).

\subsection{Density collapse at high temperature}
\la{sec:exp-high}

\begin{figure}[t]  % !htb]
\centering
\subfigure{
\includegraphics[scale=0.25]{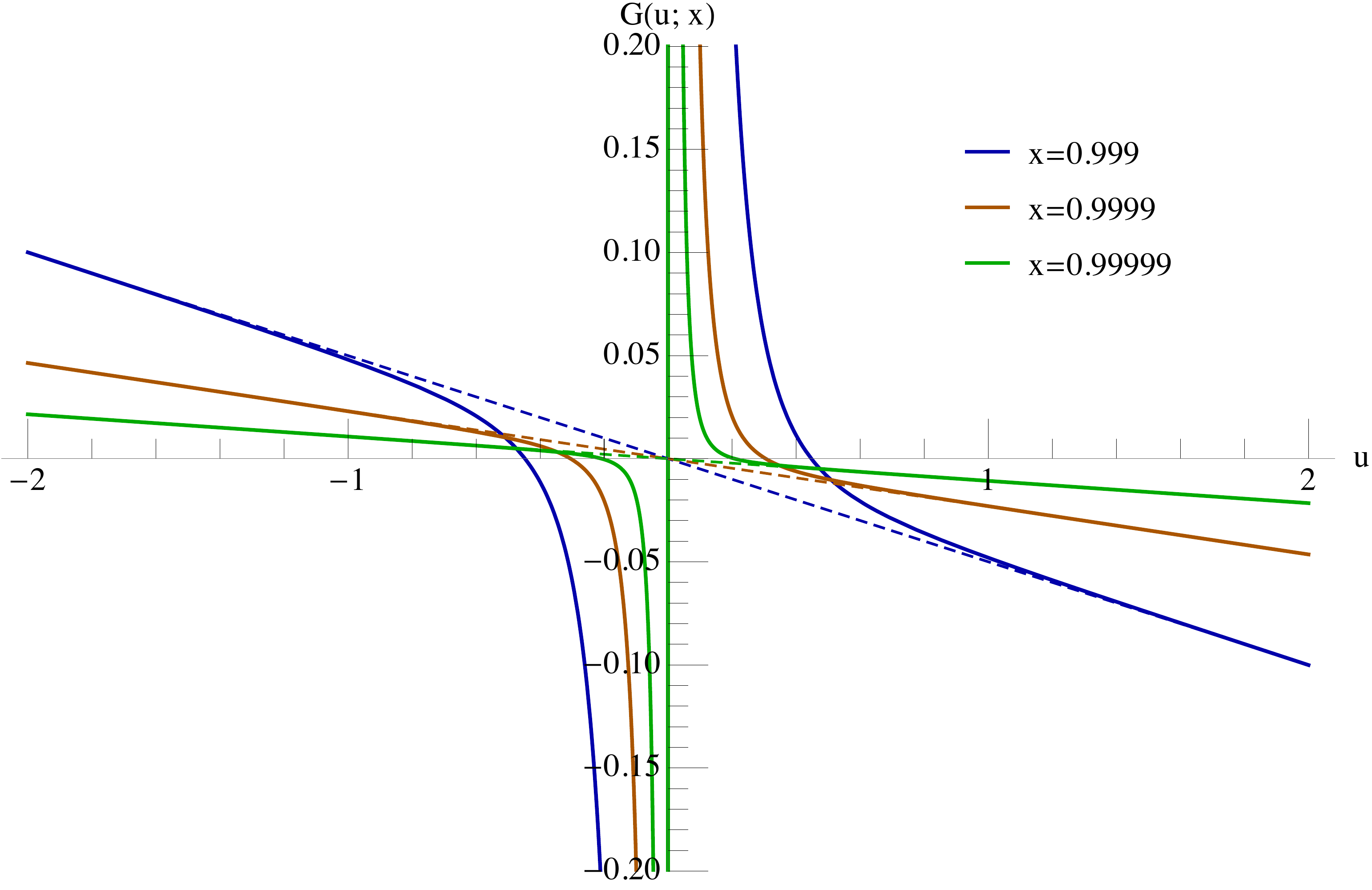}}
\subfigure{
\includegraphics[scale=0.25]{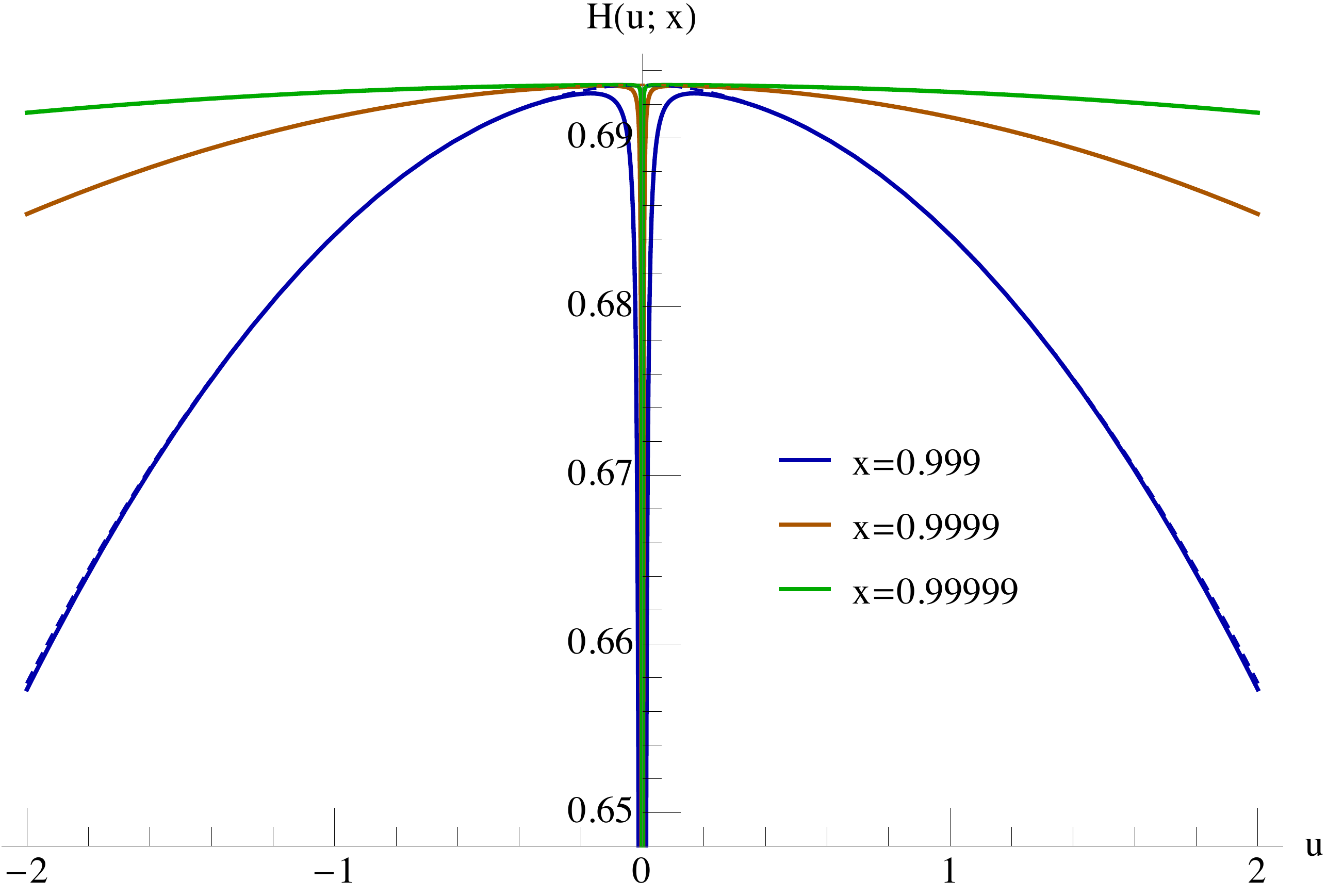}}
\caption[] {
Detailed structure of some relevant integral kernels. {\bf Left:} This panel shows 
the evaluation of the function $G$ defined in (\ref{3.10}) and evaluated with $\kappa=1$. The plot 
shows that the kernel is composed of a linear background plus a singular part which is localized
in a region of width $\sim (1-x)^{2/3}$ and approximates, as a distribution, a $\delta'$ contribution.
{\bf Right:} This panel shows 
the evaluation of the function $H$ defined in (\ref{3.20}). Similar to the left panel, we identify in the 
$x\to 1$ limit a quadratic background plus a narrow $\delta$ like contribution fully discussed in the text.
\label{fig:delta}}
\end{figure}

The expansion in the high temperature regime $x\to 1$ is more complicated and 
cannot be obtained from the formalism of Section (\ref{sec:basic}) because all $\rho_{n}$ have a non
trivial limit. Nevertheless, we can check consistency of the quadratic density (\ref{2.13})
by studying the $x\to 1$ limit of the integral equation (\ref{2.2}). This is non trivial due to the $x$
dependence of $\vartheta_{0}(x)$. Analysis of the numerical data computed in Section (\ref{sec:basic}) 
suggest that 
\be
\la{3.7}
\vartheta_{0} = \kappa\,  (1-x)^{1/3}+\dotsb.
\ee
Actually, this Ansatz may be self-consistently 
checked in the following together with the determination of the amplitude $\kappa$.
To this aim, the density can be rescaled
\be
\rho(\vartheta) = \frac{1}{\vartheta_{0}}\widetilde\rho(\vartheta/\vartheta_{0}),\qquad
\int_{-1}^{1}du\, \widetilde\rho(u)=1,
\ee
and the integral equation (\ref{2.2}) can be written in the new variables
\be
\la{3.9}
\int_{-1}^{1}du'\bigg[
\cot\left(\vartheta_{0}\,\frac{u-u'}{2}\right)-\frac{4\,x\,(1+x^{2})\,\sin(\vartheta_{0}(u-u')}{x^{4}+1
-2\,x^{2}\,\cos(2\,\vartheta_{0}\,(u-u'))}
\bigg]\,\widetilde\rho(u') = 0.
\ee
Let us  denote the kernel in the integral as $G(u; x)$, it is useful to plot it as a function of $u$ 
at various $x\to 1$ with the substitution $\vartheta_{0} \to  \kappa\,(1-x)^{1/3}$. 
This is shown in the left panel of Fig.~(\ref{fig:delta})  where $\kappa=1$
to see what is going on. As $x\to 1$, the kernel splits into
the sum of a linear {\em background} plus a $\delta'(u)$ term which is localized in a  region 
of  width $\sim (1-x)^{2/3}$. The background part comes from the naive $x\to 1$ expansion of
\be
\la{3.10}
G(u,x) = \cot\left(\kappa(1-x)^{1/3}\frac{u}{2}\right)-\frac{4\,x\,(1+x^{2})\,
\sin(\kappa\,(1-x)^{1/3}\,u)}{x^{4}+1
-2\,x^{2}\,\cos(2\,\kappa\,(1-x)^{1/3}\,u)}.
\ee
This gives
\be
\la{3.11}
G(u,x) = -\frac{1}{2}\,\kappa\,u\,(1-x)^{1/3}+\dotsb,
\ee
which may be used for $|u|\gg (1-x)^{2/3}$.
The second contribution comes from the integral (\ref{3.9}) after a zooming associated with 
$u = (1-x)^{2/3}\,\xi$. At leading order, we get \footnote{At leading order, the integration region of 
$\xi$ is symmetric and we can drop all odd contributions, some of which requires a principal value
definition.}
\begin{align}
\la{3.12}
\int_{-1}^{1}du' \,G(u-u'; x)\,\widetilde\rho(u') &= -2\,(1-x)^{1/3}\,
\widetilde\rho\,'(u)\,\int_{-\infty}^{\infty}
d\xi\,\frac{1}{\kappa\,(1+\kappa^{2}\,\xi^{2})}+\dotsb\notag \\
& = -\frac{2\,\pi}{\kappa^{2}}\,(1-x)^{1/3}\,
\widetilde\rho\,'(u)+\dotsb,
\end{align}
which has indeed the form of a $\delta'(u)$ contribution in the kernel.
Consistency of the power $1/3$ in the $1-x$ factor in (\ref{3.11}) and (\ref{3.12}) is important to get a 
non trivial result and checks  our scaling hypothesis. In summary,
at this order in the $x\to 1$ expansion, 
the integral equation becomes simply
\be
 -\frac{1}{2}\,\kappa\,u-\frac{2\,\pi}{\kappa^{2}}\,
\widetilde\rho\,'(u)=0.
\ee
This gives both the  quadratic density and the constant $\kappa$ in $\vartheta_{0}(x)$, see (\ref{3.7}),
\be
\widetilde\rho(u) = \frac{3}{4}\,(1-u^{2}),\qquad 
\kappa = (6\,\pi)^{1/3},
\ee
in agreement with (\ref{2.11}) and (\ref{2.13}).

\subsection{The transition order parameter}
\la{sec:exp-logZ}

Further consistency checks of the derived aymptotic densities come from the analysis of the 
large $N$ behaviour of $\log Z_{0}$, {\em i.e.} the free energy up to trivial factors. 
The function $F_{2}(x)$
appearing as the leading term in the large $N$ expansion  
\be
\log Z_{0} = N^{2}\,F_{2}(x) + \mc O(N\log N),
\ee
can be computed from the density $\rho(\vartheta; x)$ as  the double integral 
\be
\la{3.16}
F_{2}(x) = \frac{1}{2}\int_{-\vartheta_{0}(x)}^{\vartheta_{0}(x)} d\vartheta\,
d\vartheta'\,\log\bigg[4\,\sin^{2}\left(\frac{\vartheta-\vartheta'}{2}\right) \,
\frac{1+x^{2}+2\,x\,\cos(\vartheta-\vartheta')}{1+x^{2}-2\,x\,\cos(\vartheta-\vartheta')}
\bigg]\,\rho(\vartheta)\,\rho(\vartheta').
\ee
As we mentioned in the introduction, 
the function $F_{2}(x)$ can be regarded as an order parameter for the Hagedorn transition. 
It is zero for $0<x<x_{\rm H}$ and increases monotonically for $x>x_{\rm H}$. The maximum 
value is attained at $x=1$ and is $F_{2}(1) = \log 2$. This follows from the exact relation
 \cite{Curtright:2017pfq}
\be
Z_{0}(x\to 1) = \left(\frac{2}{1-x}\right)^{N-1}\frac{R_{N}}{N!},
\ee
where $R_{N}$ is the number of labeled Eulerian digraphs with $N$ nodes. \footnote{Basic information
about this 
sequence may be found at the OEIS link \url{http://oeis.org/A007080}.} 
The asymptotic behaviour of $R_{N}$ has been recently computed in   \cite{Curtright:2017pfq}
and reads
\be
R_{N}\stackrel{N\to \infty}{\sim} \left(\frac{2^{N}}{\sqrt{\pi N}}\right)^{N-1}\,e^{-1/4}\,\sqrt{N}\,
\left[1+\frac{3}{16N}+\mc O(N^{-2})\right],
\ee
from which we get the term $N^{2}\,\log 2$ in $\log Z_{0}$. 

Near the Hagedorn transition, we can evaluate $F_{2}(x)$ using the distribution (\ref{2.12}).
Direct expansion around $x=x_{\rm H}$ gives a leading linear behaviour 
\be
\la{3.19}
F_{2}(x) = \text{c}\,(x-x_{\rm H})+\dotsb,
\ee
where $\text{c}$ is a constant that is obtained from a rather involved finite double integral. It can be safely 
extracted from the ratio $F_{2}(x_{\rm H}+\eps)/\eps$ as $\eps\to 0$. Using $\eps = 0-10^{-3}$
and a fit of the form $a+b\,\sqrt\eps$, we reproduce the numerical data very well with $\text{c}=1/2$
with a precision of one part in $10^{6}$. For this reason, we assume that this value of $\text{c}$ is exact.
A rather small range of values of $\eps$ is needed suggesting again that the expansion around the Hagedorn 
temperature is only asymptotic. This is quite reasonable in this case because  $F_{2}(x)$ is certainly
not analytic at $x_{\rm H}$ -- it is zero below the Hagedorn temperature and non zero above it. 
The linear behaviour 
(\ref{3.19}) was originally predicted in \cite{Thorn:2015bia}.

\begin{figure}[t]  % !htb]
\centering
\includegraphics[scale=0.5]{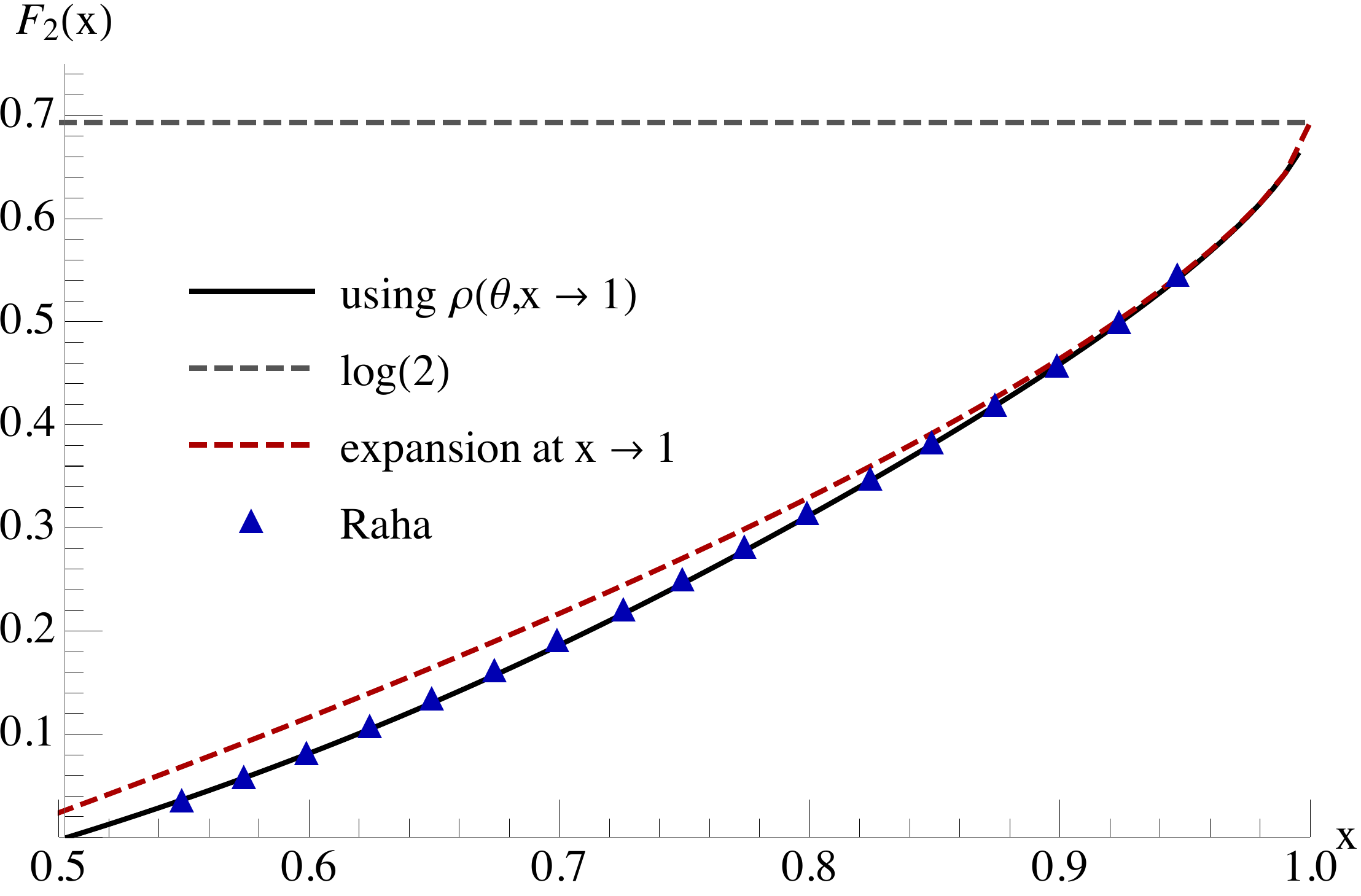}\ \ \ 
\caption[] {
Evaluation of the order parameter $F_{2}(x)$. The black curve is the result of the evaluation 
of (\ref{3.16}) using the asymptotic quadratic density in (\ref{2.13}). Blue triangles are 
exact numerical data points taken from \cite{Raha:2017jgv}. Finally, the brown dashed curve is  the 
analytical approximation in (\ref{1.7}).
\label{fig:F}}
\end{figure}

The computation of the leading correction to $F_{2}(x)$ for $x\to 1$ is more tricky and, again, 
it is again important to analyze in details the structure of the rescaled kernel 
with the leading order expression for $\vartheta_{0}$, {\em i.e.}
\be
\la{3.20}
H(u; x) = \frac{\vartheta_{0}^{2}}{2}\,
\log\bigg[4\,\sin^{2}\left(\vartheta_{0}\frac{u}{2}\right) \,
\frac{1+x^{2}+2\,x\,\cos(\vartheta_{0}\,u)}
{1+x^{2}-2\,x\,\cos(\vartheta_{0}\,u)}
\bigg],\qquad \vartheta_{0} = \kappa\,(1-x)^{1/3}.
\ee
A plot of $H(u,x)$ as a function of $u$ with $x\to 1$ is shown in the right panel
of Fig.~(\ref{fig:delta}). There is a naive quadratic contribution that comes from the 
direct expansion of $H$,
\be
H(u; x) = \log 2-\frac{(3\,\pi)^{2/3}}{4\cdot 2^{1/3}}\,u^{2}\,(1-x)^{2/3}+\dotsb.
\ee
Integrating over $\vartheta$, $\vartheta'$ in (\ref{3.16}), this gives a first contribution to $F_{2}(x)$
\be
\la{3.22}
F_{2}^{(a)}(x) = \log 2 -\frac{(6\,\pi)^{2/3}}{20}\,(1-x)^{2/3}.
\ee
A second contribution comes from zooming in the region $u-u'\sim (1-x)^{2/3}$ as in the previous
section. This gives a second $\delta(u-u')$-like  contribution leading to 
\be
\la{3.23}
F_{2}^{(b)}(x) = \frac{3}{10}(1-x)^{2/3}
\int_{-\infty}^{\infty}d\xi \log\left(\frac{\kappa^{2}\xi^{2}}{1+\kappa^{2}\xi^{2}}\right)
= -\frac{3\pi}{5\kappa}\,(1-x)^{2/3}.
\ee
Summing (\ref{3.22}) and (\ref{3.23}), we obtain the expansion (\ref{1.7}). 
In Fig.~(\ref{fig:F}), we show the evaluation of (\ref{3.16}) using the leading order
density (\ref{2.13}) with $\vartheta_{0}$ as in (\ref{2.11}). We also show the approximation (\ref{1.7})
as well as the exact numerical data points obtained in \cite{Raha:2017jgv}. The agreement is 
remarkable despite the fact that we used the asymptotic density valid for $x\to 1$.  This shows that 
$F_{2}(x)$ appears to be  little dependent on the fine structure of the density itself. This is further confirmed
by the fact that evaluation of $F_{2}(x)$ with the $x\to x_{\rm H}$ density or with the $x\to 1$ one
are practically indistinguishable up to $x\simeq 0.9$.

\section{Conclusions}

In this paper we have considered the large $N$ thermodynamics of a simple $SU(N)$
string bit model devised to capture   the tensionless limit of the associated 
string. The model lives in the color singlet sector and involves a projection implemented by a suitable
group average, {\em i.e.} integration over $U\in SU(N)$.
Dominant configurations are characterized by a non trivial density $\rho(\vartheta; T)$
of the $U$ invariant phases $\vartheta_{1}, \dots, \vartheta_{N}$.
We have analyzed the model in the gapped phase, {\em i.e.}
for temperatures above the Hagedorn transition $T>T_{\rm H}$ where $\rho(\vartheta; T)$
is non zero only in the interval $|\vartheta|\le \vartheta_{0}(T)<\pi$.
By means of numerical and analytical tools, we
have discussed in some details the 
temperature dependence of the phase density $\rho(\vartheta; T)$ including 
the gap endpoint $\vartheta_{0}(T)$. 
Our results  provide quantitative information about  the crossover from the low to high temperature
phases in the considered model. 
It remains an open question to understand precisely which changes occur 
in models with more bits and if $1/N$ corrections are taken into account to 
resolve the phase transition. The corrections we 
found at $N=\infty$ contains non trivial power exponents, see (\ref{1.5}) and (\ref{1.6}). 
In particular, the phase density support $[-\vartheta_{0},\vartheta_{0}]$ opens a gap in the $\vartheta$
distribution of width
$2\,(\pi-\vartheta_{0})\sim (T-T_{\rm H})^{1/4}$ near the Hagedorn transition. Besides, the support 
collapses with $\vartheta_{0}\sim T^{-1/3}$
for $T\gg T_{\rm H}$. It could be  interesting to understand these relations 
in the context of a finite but large $N$  double scaling limit as  in the 
Hagedorn transition for IIB string theory in an anti-de 
Sitter spacetime \cite{Liu:2004vy,AlvarezGaume:2005fv}.

\bibliography{BT-Biblio}
\bibliographystyle{JHEP}

\end{document}